\documentclass[proceedings]{JHEP3} 
\PrHEP{ hep2001}
\usepackage{epsfig,multicol}            
\newbox\mybox
\newcommand\fverb{\setbox\mybox=\hbox\bgroup\verb}
\newcommand\fverbdo{\egroup\medskip\noindent\fbox{\unhbox\mybox}\ }
\newcommand\fverbit{\egroup\item[\fbox{\unhbox\mybox}]}
\def\pr{\partial}
\def\prd{\partial \cdot}
\def\btd{\bigtriangledown}

\def\a{\alpha}
\def\b{\beta}

\def\G{\Gamma}

\def\d{\delta}
\def\e{\epsilon}

\def\L{\Lambda}
\def\m\mu
\def\n{\nu}

\def\vf{\varphi}

\def\dsll{\not {\! \pr}}
\def\xisl{\not {\! \xi}}
\def\psisl{\not {\! \! \psi}}
\def\nablasl{\not {\! \! \nabla}}
\def\esl{\not {\! \epsilon}}
\def\ssl{\not {\! \cal S}}
\def\12{\frac{1}{2}}
\def\fr{\frac}
\def\pr{\partial}
\def\prd{\partial \cdot}
\def\btd{\bigtriangledown}

\def\a{\alpha}
\def\b{\beta}

\def\G{\Gamma}

\def\d{\delta}
\def\e{\epsilon}

\def\L{\Lambda}
\def\m\mu
\def\n{\nu}

\def\vf{\varphi}

\def\dsll{\not {\! \pr}}
\def\psisl{\not {\! \! \psi}}
\def\esl{\not {\! \epsilon}}
\def\ssl{\not {\! \cal S}}
\title{On the geometry of higher-spin gauge fields}
\author{{Dario Francia\footnote{Present address: Dipartimento di Fisica, Universit\`a di Roma Tre and INFN, Sezione
di Roma Tre, Via della Vasca Navale 84, 00146 Roma \ ITALY.} \ and
\ Augusto Sagnotti}
\\
    Dipartimento di Fisica, Univ. Roma ``Tor Vergata'' and I.N.F.N., Sezione di Roma ``Tor Vergata''\\
    Via della Ricerca Scientifica, 1 \ 00133 Roma \ ITALY\\
    E-mail: \email{sagnotti@roma2.infn.it}}
\conference{\raisebox{4pt}{\parbox{11cm}%
{\raggedleft \small 27th Johns
Hopkins Workshop on Current Problems in Particle Theory:\\
Symmetries and Mysteries of M Theory}}}
\abstract{We review a recent construction of the free field
equations for totally symmetric tensors and tensor-spinors that
exhibits the corresponding linearized geometry. These equations
are {\it not local} for all spins $>2$, involve unconstrained
fields and gauge parameters, rest on the curvatures introduced
long ago by de Wit and Freedman, and reduce to the local
(Fang-)Fronsdal form upon partial gauge fixing. We also describe
how the higher-spin geometry is realized in free String Field
Theory, and how the gauge fixing to the light cone can be
effected. Finally, we review the essential features of local
compensator forms for the higher-spin bosonic and fermionic
equations with the same unconstrained gauge symmetry.}
\begin{document} \vskip 5pt
\noindent \emph{Based on talks presented by A.S. at the The Third
International Sakharov Conference on Physics, Moscow, June 24-29
2000, at the XV SIGRAV Conference on General Relativity and
Gravitational Physics, Villa Mondragone (Rome), September 9-12
2002 and at the Leuven Workshop on ``The quantum structure of
space time'', September 13-19 2002, and on the third Lecture
delivered by A.S. at the Cargese Summer Institute on ``Progress in
String, Field and Particle Theory'', June 25-July 11 2002. Updated
version based on talks presented by A.S. at the Johns Hopkins
Workshop, Goteborg, Aug. 24-26 \ 2003, at the Gunnar Nordstr\"om
Symposium, Helsinki, Aug. 27-30 \ 2003, at the INFN-BO11 Workshop,
Bologna, Sept. 16-17, 2003 and at the INFN-TS11 Worskhop, Perugia, Dec.
19-20, 2003.}

\section{Introduction}

This article reviews the results of \cite{fs,sagtsu} on the geometry
underlying the free equations for symmetric tensors of arbitrary
spin $s$ and on its role in corresponding equations for fermionic
tensors of spin $s+1/2$. It is meant as a contribution to the
Conference Proceedings of the Sakharov, Leuven and SIGRAV
Meetings, and to the Proceedings of the 2002 Cargese Summer
Institute. The SIGRAV contribution was accompanied by a less
technical lecture at the companion SIGRAV School, essentially
based on \cite{moriond}, while only one of the three Cargese
lectures was actually devoted to this topic. However, since the
other two dealt with results reviewed in detail in \cite{prep}, it
will suffice to confine the attention to this topic. This will
also provide an opportunity to supplement the results in \cite{fs}
with additional remarks on their role in String Theory and on the
issue of gauge fixing \footnote{We are grateful to A. Segal for
his stimulating remarks on this issue.}. The last two sections, added
in this new version, review the results in \cite{sagtsu} and
present local compensator forms for the higher-spin bosonic and
fermionic equations with the same unconstrained gauge symmetry.

The starting point for our discussion are the elegant equations
extracted long ago by Fronsdal \cite{fronsdal} from the massive
Singh-Hagen \cite{singhag} actions,
\begin{equation}
{\cal F} \ \equiv \ \Box \, \phi \ -\  \partial \, \partial \cdot
\phi \ +  \ \partial^2 \, \phi^{\, \prime} \ = \ 0 \ ,
\label{fronsdeq}
\end{equation}
that for spin one and two reduce to the Maxwell equation for the
vector potential $A_\mu$ and to the linearized Einstein equation
for the metric fluctuation $h_{\mu\nu}$. Here, as will often be
the case in the following, primes (or bracketed suffixes) denote
traces, while all indices carried by the symmetric tensors
$\phi_{\mu_1 \dots \mu_s}$ and $\eta_{\mu\nu}$ or by derivatives
are left implicit. In this shorthand notation, where all terms are
meant to be totally symmetrized, one need only get accustomed to a
few rules, some notable examples for a $D$-dimensional space time
being
\begin{eqnarray}
&& \left( \partial^{\; p} \ \phi  \right)^{\; \prime} \ = \ \Box \
\partial^{\; p-2} \
\phi \ + \ 2 \, \partial^{\; p-1} \  \partial \cdot \phi \ + \
\partial^{\; p} \
\phi {\;'} \ , \\
&& \partial^{\; p} \ \partial^{\; q} \ = \ \left( {p+q} \atop p
\right) \
\partial^{\; p+q} \ ,
\\
&& \partial \cdot  \left( \partial^{\; p} \ \phi \right) \ = \
\Box \
\partial^{\; p-1} \ \phi \ + \
\partial^{\; p} \ \partial \cdot \phi \ ,  \\
&& \partial \cdot  \eta^{\;k} \ = \ \partial \, \eta^{\;k-1} \ , \\
&& \left( \eta^k \, T_{(s)} \,  \right)^\prime \ = \ k \, \left[
\, D \, + \, 2(s+k-1) \,  \right]\, \eta^{k-1} \, T_{(s)} \ + \
\eta^k \,  T_{(s)}^\prime \ , \label{etak}
\end{eqnarray}
where $T_{(s)}$ denotes a symmetric rank-$s$ tensor, but the
general case can then be dealt with in a rather efficient and
compact fashion.

At a closer look, however, for spin $s > 2$ eq. (\ref{fronsdeq})
appears somewhat less natural than its first and most familiar
instances,
\begin{eqnarray}
& & \Box A_\mu \ - \ \partial_\mu \, \partial \cdot A \ = \ 0 \ , \\
& & \Box h_{\mu\nu} \ - \ \partial_\mu \, \partial \cdot h_\nu \ -
\
\partial_\nu \, \partial \cdot h_\mu \ + \
\partial_\mu \partial_\nu h^{\; \prime} \ = \ 0 \ ,
\end{eqnarray}
since only these involve {\it all} available lower spin constructs
built from divergences and traces of the gauge field. This
peculiarity has in fact an important consequence: only for spin
one and two does the gauge transformation
\begin{equation}
\delta \; \phi \ = \ \partial \; \Lambda \label{gaugetransf}
\end{equation}
leave the Fronsdal operator ${\cal F}$ invariant, while in general
\begin{equation}
\delta \; {\cal F} \ = \ 3 \ \partial^{\; 3} \, \Lambda^{\;
\prime} \ ,
\end{equation}
and as a result for spin $s \geq 3$ eq. (\ref{fronsdeq}) requires
a {\it traceless} gauge parameter $\Lambda$. This constraint plays
a key role in the Fronsdal formulation \cite{fronsdal}, together
with an additional one, effective from $s=4$ onward, on the gauge
field, whose {\it double trace} $\phi''$ is also required to
vanish.

We can now review how one can actually forego both constraints, at
the price of allowing {\it non-local} terms both in the free field
equations and in the corresponding Lagrangians, but with the clear
advantage of eliciting the geometry that underlies them. This
geometry would be naturally expected to play a role in the
Vasiliev equations \cite{fvas,vaseq}, but to date we have not gone
far enough to make definite statements to this effect. However, it
is a fact that String Field Theory \cite{bpsz,witten}, in the most
symmetric formulation of its free field equations, deduced in
\cite{witten} from the BRST analysis of \cite{ko}, {\it does not}
involve trace conditions on the gauge fields or on the
corresponding gauge parameters. Rather, as we shall see, it
embodies very nicely the higher-spin geometric equations that we
are about to review.

Let us therefore begin by tracing the origin of the double trace
constraint on the gauge field. To this end, it is important to
recall that, already at spin two, the naive linearized Einstein
equation
\begin{equation}
R_{\mu\nu}^{(lin)} \ = \ 0
\end{equation}
does not follow directly from the Einstein-Hilbert action
principle, whose variation actually gives
\begin{equation}
\delta {\cal L} \sim \delta h_{\mu\nu} \, \left[ \,
R_{\mu\nu}^{(lin)} \ - \ \frac{1}{2} \, \eta_{\mu\nu} \, R^{(lin)}
\, \right]  \ . \label{einsteintens}
\end{equation}
There is a clear logic behind this result, however, since the
linearized Einstein tensor
\begin{equation}
G_{\mu\nu}^{(lin)} = R_{\mu\nu}^{(lin)} \ - \ \frac{1}{2} \,
\eta_{\mu\nu} \, R^{(lin)}
\end{equation}
has the special virtue of being divergence free, and this ensures
that $\delta L$ vanishes, up to a total derivative,
 if $\delta h^{\mu\nu}$ corresponds to a gauge
transformation. In other words, the gauge invariance of $L$ rests
on two ingredients: the gauge invariance of the field equation and
a suitable Bianchi identity. This pattern extends to all
higher-spin symmetric tensors, whose Bianchi identities
\begin{equation}
\partial \cdot {\cal F} \  - \ \frac{1}{2} \ \partial \, {\cal F}\;' \ = \ - \
\frac{3}{2} \ \partial^{\;3}\, \phi^{''} \ , \label{bianchifron}
\end{equation}
however, contain a sort of {\it classical anomaly} that involves
their double traces. As a result, for spin $> 2$ the Einstein-like
tensors
\begin{equation}
{\cal G} = {\cal F} \ - \ \frac{1}{2} \, \eta \, {\cal F}'
\end{equation}
can only be compatible with gauge invariant Lagrangians such that
\begin{equation}
\delta {\cal L} \sim \delta \phi \left[ \, {\cal F} \ - \
\frac{1}{2}\, \eta \,  {\cal F}' \, \right] \label{deltaL}
\end{equation}
if the gauge parameters $\Lambda$ of eq. (\ref{gaugetransf}) are
{\it traceless} and the gauge fields $\phi$ are {\it doubly
traceless}. Using the combinatoric rules listed above one can
indeed show that, up to a total derivative, the gauge
transformation (\ref{gaugetransf}) turns eq. (\ref{deltaL}) into
\begin{equation}
\delta {\cal L} \sim \Lambda \left[ \, \partial \cdot {\cal F} \ -
\ \frac{1}{2}\,
\partial \, {\cal F}'
\ - \ \frac{1}{2} \, \eta \, \partial \cdot {\cal F}' \, \right] \
,
\end{equation}
that can only vanish in general if both restrictions are enforced.
Notice that the Bianchi identity eliminates the first two terms,
while the last does not vanish identically. Rather, via the $\eta$
tensors, it places again on the gauge parameter the same
restriction that ensures the gauge invariance of the Fronsdal
operator ${\cal F}$. To reiterate, the restriction to traceless
gauge parameters is instrumental in the gauge invariance of the
Fronsdal equations, while the restriction to doubly traceless
fields ensures that the corresponding action principles be
compatible with the gauge symmetry.

\section{Gauge-invariant equations for unconstrained bosons}

In order to forego these restrictions, one can follow an iterative
procedure that, as in \cite{fs}, can be conveniently motivated in
the relatively simple spin-3 case, where
\begin{equation}
\delta {\cal F}_{\mu_1\mu_2\mu_3} \ = \ 3\,  \partial_{\mu_1} \,
\partial_{\mu_2} \, \partial_{\mu_3} \, \Lambda' \ ,
\label{fronsdvar}
\end{equation}
since one can find rather simply other {\it non-local} constructs
of ${\cal F}$ that vary as in (\ref{fronsdvar}), to wit
\begin{eqnarray}
&& \frac{1}{3 \, \Box} \, \left[ \, \partial_{\mu_1}
\partial_{\mu_2} {\cal F}{\;'}_{\mu_3} +   \partial_{\mu_2}
\partial_{\mu_3} {\cal F}{\;'}_{\mu_1} +
\partial_{\mu_3}
\partial_{\mu_1} {\cal F}{\;'}_{\mu_2} \,  \right] \ , \nonumber \\
&& \frac{1}{3 \, \Box} \, \left[ \, \partial_{\mu_1} \partial
\cdot {\cal F}_{\mu_2\mu_3} +
\partial_{\mu_2} \partial \cdot {\cal F}_{\mu_3\mu_1} +
\partial_{\mu_3} \partial \cdot {\cal F}_{\mu_1\mu_2} \, \right] \ , \nonumber \\
&& \frac{1}{ \Box^{\;2}} \ \partial_{\mu_1} \partial_{\mu_2}
\partial_{\mu_3} \, \partial \cdot {\cal F}{\;'} \ .
\end{eqnarray}
The first two expressions actually coincide, as can be seen making
use of the Bianchi identity (\ref{bianchifron}), and as a result
one is led to two apparently distinct forms for a non-local fully
gauge invariant field equation:
\begin{eqnarray}
&& {\cal F}_{\mu_1\mu_3\mu_3} \ - \ \frac{1}{3 \, \Box} \ \left[
\; \partial_{\mu_1} \partial_{\mu_2} {\cal F}{\;'}_{\mu_3} +
\partial_{\mu_2} \partial_{\mu_3} {\cal F}{\;'}_{\mu_1} +
\partial_{\mu_3}
\partial_{\mu_1} {\cal F}{\;'}_{\mu_2}  \; \right] = 0 \ , \label{Fn} \\
&& {\cal F}_{\mu_1\mu_2\mu_3} \ - \ \frac{1}{ \Box^{\;2}} \
\partial_{\mu_1} \partial_{\mu_2}
\partial_{\mu_3} \partial \cdot {\cal F}{\;'} = 0 \ .
\label{spinthreeh}
\end{eqnarray}
These can be actually turned into one another, once they are
combined with their traces, but the second form is clearly
somewhat simpler, since its rests on the addition of the spin-0
construct $ \partial \cdot {\cal F}{\;'}$.

For all higher spins, one can arrive at the proper analogue of
(\ref{Fn}) via a sequence of pseudo-differential operators,
defined recursively as
\begin{equation}
{\cal F}^{(n+1)} \ = \ {\cal F}^{(n)} \ + \ \frac{1}{(n+1) (2 n +
1)} \ \frac{\partial^{\;2}}{\Box} \, {{\cal F}^{(n)}}\;' \ - \
\frac{1}{n+1} \ \frac{\partial}{\Box} \
\partial \cdot  {\cal F}^{(n)} \ , \label{recursion}
\end{equation}
where ${\cal F}^{(1)}={\cal F}$, whose gauge transformations,
\begin{equation}
\delta {\cal F}^{(n)} \ = \ \left( 2 n + 1 \right) \ \frac{
\partial^{\; 2 n + 1}} {\Box^{\; n-1}} \ \Lambda^{[n]} \ ,
\end{equation}
involve by construction higher traces of the gauge parameter.
Since the $n$-th trace $\Lambda^{[n]}$ is only available for spins
$s \geq 2n+1$, after a certain number of iterations one is bound
to end up with a fully gauge invariant, albeit non-local, kinetic
operator.

A similar inductive argument determines the Bianchi identities for
the ${\cal F}^{(n)}$,
\begin{equation}
\partial \cdot {\cal F}^{(n)} \ - \ \frac{1}{2n} \
\partial {{\cal F}^{(n)}}{\; '} \ = \ - \
\left( 1 + \frac{1}{2n}  \right) \ \frac{\partial^{\;
2n+1}}{\Box^{\; n-1}} \ \phi^{[n+1]}  \ , \label{bianchin}
\end{equation}
where the ``anomalous'' contribution depends on the $(n+1)$-th
trace $\phi^{[n+1]}$ of the gauge field, and thus vanishes for all
spins $s < 2n+2$. Summarizing, these results imply that, after a
certain number of iterations, one can obtain {\it non-local}
Lagrangians and field equations whose gauge invariance places no
restrictions on the gauge fields or on the corresponding gauge
parameters. Eq. (\ref{bianchin}) also implies similar relations
for the higher traces of the ${\cal F}^{(n)}$,
\begin{equation}
\partial \cdot {\cal F}^{(n)\, [k]} \ - \ \frac{1}{2(n-k)} \
\partial {{\cal F}^{(n)\, [k+1]}} \ = \
0 \ , \qquad ( k \leq n-1) \label{bianchink}
\end{equation}
here written for $n$ large enough so that the ``anomaly'' on the
{\it r.h.s.} of (\ref{bianchin}) vanishes identically. Notice that
for odd spin $s=2n-1$ the second term vanishes for the last trace,
so that
\begin{equation}
\partial \cdot {\cal F}^{(n)\, [n-1]} \ = \ 0 \ . \label{lastodd}
\end{equation}

These generalized Bianchi identities suffice to define for all
spin-$s$ fields fully gauge invariant analogues of the Einstein
tensor,
\begin{equation}
{\cal G}^{(n)} \ = \ \sum_{p \leq n} \ \frac{(-1)^p}{2^p \ p! \
\left( {n \atop p} \right)} \ \eta^p \ {\cal F}^{(n)\, [p]} \ ,
\end{equation}
obtained combining the proper ${\cal F}^{(n)}$ with their multiple
traces that, for $n$ large enough, have vanishing divergences like
their spin-2 counterpart. This is attained directly by the
subtractions for all even spins, while for odd spins the last term
thus generated vanishes on account of (\ref{lastodd}). From the
${\cal G}^{(n)}$ tensors, integrating
\begin{equation}
\delta \, {\cal L} \ \sim \ \delta \phi \ {\cal G}^{(n)} \ ,
\end{equation}
one can then recover the corresponding gauge invariant
Lagrangians.

In order to write the fully gauge invariant equations in a compact
form, it is convenient to resort to the trick of \cite{bargtod},
contracting the gauge fields $\phi_{\mu_1 \cdots \mu_s}$ with a
vector $\xi$. It is then simple to show that traces and
divergences of the resulting expressions,
\begin{equation}
\hat{\Phi}(x,\xi) \ = \ \frac{1}{s!}\ \xi^{\mu_1} \cdots
\xi^{\mu_s} \ \phi_{\mu_1 \cdots \mu_s} \ , \label{phixi}
\end{equation}
can be recovered applying the differential operators $\partial_\xi
\cdot \partial_\xi$ and $\partial_\xi \cdot \partial$, where
$\partial_\xi$ denotes a derivative with respect to $\xi$.
Starting from the Fronsdal term
\begin{equation}
\hat{\cal F}(\hat{\Phi}) = \left[ \ \Box \ - \ \xi \cdot \partial
\ \ \partial \cdot
\partial_\xi \ + \ (\xi \cdot \partial)^2 \ \partial_\xi \cdot \partial_\xi
  \ \right]
\hat{\Phi} \ , \label{fronsdalxi}
\end{equation}
the result of the successive iterations
can then be written in the compact form\\
\begin{equation}
  \prod_{k=0}^{n-1} \left[\,  1 \, + \,  \frac{(\xi \cdot \partial)^2}
{(k+1)(2k+1)} \ \frac{ \partial_\xi \cdot \partial_\xi}{\Box} \
 \,
- \, \frac{\xi \cdot
\partial}{k+1} \ \frac{ \partial_\xi \cdot \ \partial}{\Box} \,
 \  \right]  \, \hat{\cal F}(\hat{\Phi}) \, = \, 0 \  , \label{eqnonlocgen}
\end{equation}
where for spin $s$ a fully gauge invariant operator is reached
after $\left[ \frac{s+1}{2} \right]$ iterations. Expanding this
expression and combining it with its trace it is then possible to
show that the field equations can always be reduced to the form
\begin{equation}
{\cal F} \ =  \ \partial^{\; 3} \ {\cal H} \ , \label{fd3h}
\end{equation}
that generalizes eq. (\ref{spinthreeh}). Under a gauge
transformation, by consistency, $\delta {\cal H} = 3 \ \Lambda{\;
'}$, and therefore the local forms (\ref{fronsdalxi}) can be
recovered from (\ref{eqnonlocgen}) making use of the traces of the
gauge parameters $\Lambda$ to reach the ``Fronsdal gauges'' ${\cal
H} = 0$. The final gauge fixing to the light cone, however,
presents some subtleties, and will be discussed in a later
section.

Gauge invariant Lagrangians can be also associated to eq.
(\ref{fd3h}), as can already be seen for spin 3. They differ from
those corresponding to the ${\cal F}^{(n)}$, and have the virtue
of reducing to the local expressions of \cite{fronsdal} in the
Fronsdal gauges ${\cal H} = 0$, but the general expressions for
the corresponding Einstein-like tensors are less transparent, and
we shall thus refrain from displaying them explicitly.

\section{Gauge-invariant equations for unconstrained fermions}

One can also arrive at similar non-local geometric equations for
totally symmetric tensor-spinors $\psi_{\mu_1\cdots\mu_s}$. In
this case the local fermionic equations of \cite{fronsdal}
\begin{equation}
{\cal S} \ \equiv \ i \, \left( {\not {\! \pr}} \, \psi - \pr
\psisl \right) \ = \ 0 \label{fangfroneq}
\end{equation}
are gauge invariant under
\begin{equation}
\delta \psi \ = \ \pr \, \e
\end{equation}
only if the gauge parameters are subject to the constraint
\begin{equation}
\esl \ = \ 0 \ . \label{fermiparf}
\end{equation}
In addition, the ${\cal S}$ operators satisfy the ``anomalous''
Bianchi identities
\begin{equation}
\prd {\cal S} \ - \ \frac{1}{2} \, \pr \ {\cal S}{\; '} \ - \
\frac{1}{2} {\not {\! \pr}} \ssl \
 = \ i \ \pr^{\; 2} \psisl\;' \ , \label{fangfronbianchi}
\end{equation}
and therefore the gauge variation of the generic Lagrangian
\begin{equation}
\d {\cal L} \ \sim\ \d \bar{\psi} \left[\, {\cal S} \ - \
\frac{1}{2} \left(\, \eta \, {\cal S}' \ + \ \gamma \ssl \,
\right) \, \right]
\end{equation}
can only vanish if eq. (\ref{fermiparf}) is supplemented by
\begin{equation}
\psisl\; ' \ = \ 0 \ , \label{fermiparf2}
\end{equation}
the fermionic analogue of the double trace condition for boson
fields.

It is convenient to notice that the fermionic operators for spin
$s+1/2$ can be formally related to the bosonic operators for spin
$s$ according to
\begin{equation}
{\cal S}_{s+1/2} \ - \ \frac{1}{2} \, \frac{\pr}{\Box}\,
{\not{\!\pr}} \, \ssl_{s+1/2} \ = \ i \ \frac{\not{\!\pr}}{\Box}
\, {\cal F}_s(\psi) \ . \label{bosefermi}
\end{equation}
This amusing link generalizes the obvious one between the Dirac
and Klein-~Gordon operators and, for the rather familiar $s=1$
case, connects the Rarita-Schwinger operator, that when combined
with its $\gamma$ trace can be reduced to the form
(\ref{fangfroneq}), to the Maxwell operator, according to
\begin{equation}
{\cal S}_{\mu} \ - \ \frac{1}{2} \, \frac{\pr_\mu\,
{\not{\!\pr}}}{\Box} \, \ssl \ = \ i \ \frac{\not{\!\pr}}{\Box} \,
\left[\ \Box \, \eta_{\mu\nu} \ - \ \pr_\mu \, \pr_\nu    \
\right]\, A^\nu \ , \label{bosefermi1}
\end{equation}
where we have restored the conventional index notation.

One can also show, by an inductive argument, that similar
relations hold between the non-local fully gauge invariant
counterparts of eqs. (\ref{fronsdeq}) and (\ref{fangfroneq}). They
link the fermionic kinetic operators ${\cal S}^{(n)}$, defined
recursively as
\begin{equation}
{\cal S}^{(n+1)} \ = \ {\cal S}^{(n)} \ + \ \frac{1}{n(2n+1)} \,
\frac{\pr^{\; 2}}{\Box} \, {\cal S}^{(n)\; '} \ - \ \frac{2}{2n+1}
\, \frac{\pr}{\Box} \, \prd {\cal S}^{(n)} \ ,
\end{equation}
and such that
\begin{equation}
\d \, {\cal S}^{(n)} \ = \ - \ 2 \, i \, n \ \frac{\pr^{\;
2n}}{\Box^{\; n-1}}\,  \esl^{\; [n-1]} \ ,
\end{equation}
to the corresponding corrected bosonic operators of the previous
section, according to
\begin{equation}
{\cal S}^{(n)}_{s+1/2} \ - \ \frac{1}{2n} \, \frac{\pr}{\Box}\,
{\not{\!\pr}} \, \ssl_{s+1/2}^{(n)} \ = \ i \
\frac{\not{\!\pr}}{\Box} \, {\cal F}^{(n)}_s(\psi) \ .
\end{equation}

This relation also determines the ``anomalous'' Bianchi identities
of the ${\cal S}^{(n)}$,
\begin{equation}
\prd {\cal S}^{(n)} \ - \ \frac{1}{2n} \, \pr \ {\cal S}^{(n)\; '}
\ - \ \frac{1}{2n}\, {\not {\! \pr}} \ssl^{(n)} \  = \ i \
\frac{\pr^{\; 2n}}{\Box^{\; n-1}} \psisl^{[n]} \ ,
\label{bianchifermi}
\end{equation}
and therefore, for $n$ large enough so that $S^{(n)}$ is gauge
invariant and eq. (\ref{bianchifermi}) is free of the
``anomalous'' term, the corrected Einstein-like operators
\begin{equation}
{\cal G}^{(n)} \ = \ {\cal S}^{(n)} \ + \ \sum_{0 < p \leq n} \
\frac{(-1)^p}{2^p \ p! \ \left( {n \atop p} \right)} \ \eta^{p-1}
\left[\  \eta \ {\cal S}^{(n)\, [p]} \ + \ \gamma  \ {\cal {\not
{\! S}}}^{(n)\, [p-1]} \ \right] \ ,
\end{equation}
whose divergences vanish on account of eqs. (\ref{bianchifermi})
and of their ($\gamma$-) traces, and from which gauge invariant
Lagrangians can be recovered. For the sake of brevity, in the
following we shall mostly restrict our attention to bosonic
higher-spin fields.

\section{Free-field geometry}

Following de Wit and Freedman \cite{dewf}, one can define
generalized connections of various orders in the derivatives for
all spin-$s$ gauge fields. This can be done by an iterative
procedure, so that, in the compact notation of the previous
section, for any field of spin $s$, after $m$ iterations one can
define
\begin{equation}
\G^{(m)} =  \frac{1}{m+1} \ \sum_{k=0}^{m}\fr{(-1)^{k}}{ \left(
{{m} \atop {k}} \right) }\ \pr^{m-k}\btd^{k} \vf \ ,
\label{gammam}
\end{equation}
where we are now using two types of derivatives for two distinct
sets of indices, $\pr$ for the $m$ symmetric indices $(\a_1 \cdots
\a_m)$ introduced by the derivatives and $\btd$ for the other $s$
symmetric ones $(\b_1 \cdots \b_s)$ originating from the gauge
field. It is simple to show, by an inductive argument, that the
gauge transformation of $\G^{(m)}$ is
\begin{equation}
\delta \, \G^{(m)} = \btd^{m+1} \L \ , \label{deltagammam}
\end{equation}
so that all $m$ indices of the first set are within the gauge
parameter. Hence,
\begin{equation}
\G^{(s-1)} = \fr{1}{s} \sum_{k=0}^{s-1}\fr{(-1)^{k}}{ \left(
{{s-1} \atop {k}} \right) }\ \pr^{s-k-1}\btd^{k} \phi \ ,
\label{connection}
\end{equation}
is the proper analogue of the Christoffel connection for a
spin-$s$ gauge field, since its gauge transformation contains a
single term. That these objects can be defined in general can be
also recognized noticing that the $(s-1)$-th derivatives of the
spin-$s$ gauge transformations of eq. (\ref{gaugetransf}) imply
that one can retrieve a composite connection $\G_{\a_1 \cdots
\a_{s-1}; \b_1 \cdots \b_s}$, such that
\begin{equation}
\delta\G_{\a_1 \cdots \a_{s-1}; \b_1 \cdots \b_s}  = \pr_{\b_1}
\cdots \pr_{\b_s} \ \L^{ \a_1 \cdots \a_{s-1}} \ ,
\label{gaugetransfgammas}
\end{equation}
inverting a linear system with $\left( {2s-1 \atop s } \right)$
unknowns. Moreover, {\it all} $\G$'s with $m >s$ are gauge
invariant, and in particular
\begin{equation}
\G^{(s)} = \fr{1}{s+1}\sum_{k=0}^{s}\fr{(-1)^{k}}{ \left( {{s}
\atop {k}} \right) }\ \pr^{s-k}\btd^{k} \phi  \label{riemann}
\end{equation}
is the proper analogue of the Riemann curvature tensor. This
generalized curvature is of the form
 ${\cal R}_{\alpha_1 \cdots \alpha_s;\beta_1 \cdots
\beta_s}$, and is totally symmetric under the interchange of any
pair of indices within each of the two sets. In addition, as shown
in \cite{dewf},
\begin{equation}
{\cal R}_{\alpha_1 \cdots \alpha_s;\beta_1 \cdots \beta_s} =
(-1)^s \, {\cal R}_{\beta_1 \cdots \beta_s;\alpha_1 \cdots
\alpha_s} \ ,
\end{equation}
and a generalized cyclic identity holds. These concepts can be
also related to an interesting generalization of the exterior
differential, whereby the familiar condition $d^2=0$ is replaced
by $d^{s+1}=0$ \cite{henn}.

There is another, perhaps more obvious way, to generate a gauge
invariant quantity from a connection $\G^{(s-1)}$ that transforms
as in (\ref{gaugetransfgammas}), a curl with respect to any of its
$\b$ indices. However, the choice of \cite{dewf} has the virtue of
simplicity, since it results automatically in a tensor with two
totally symmetric sets of indices. For instance, the symmetric
curvature of \cite{dewf} for the spin-2 case is a linear
combination of ordinary Riemann tensors, while its trace in the
first symmetric set of indices is proportional to the ordinary
Ricci tensor. If we now restrict our attention to the $\G^{(m)}$'s
with $m$ {\it even}, and for the sake of clarity let $m=2n$, eq.
(\ref{deltagammam}) implies that the total trace of $\G^{(2n)}$
over pairs of $\b$ indices, $\G^{(2n) [n]}$, is a totally
symmetric spin-$s$ tensor such that
\begin{equation}
\d \, \frac{1}{\Box^{n-1}}\ \G^{(2n)\, [n]} =
\frac{\pr^{2n+1}}{\Box^{n-1}} \, \L^{[n]} \ .
\end{equation}
Up to an overall proportionality constant, this is exactly the
gauge transformation of the ${\cal F}^{(n)}$, and in particular,
if $s=2n$, $\G^{(2n)[n]}$ is the $n$-th trace of the spin-$s$
analogue of the Riemann tensor defined above. The iterative
procedure of the previous section is thus providing a r\^ole for
the higher-spin connections of \cite{dewf}, so that the geometric
gauge-invariant equations for even spin $s=2n$ can be written in
the form
\begin{equation}
\frac{1}{\Box^{n-1}} \ {\cal R}^{[n]}{}_{;\mu_1 \cdots \mu_{2n}} =
0 \ , \label{geomeven}
\end{equation}
a natural generalization of the linearized Einstein equation.

The odd-spin case $s=2n+1$ presents a further minor subtlety, in
that the corresponding curvatures $\G^{2n+1}$ have an odd number
of $\b$ indices. The simplest option is then to take a trace over
$n$ pairs of $\b$ indices in $\G^{2n+1}$ and act with a divergence
on the remaining one. The end result for spin $s=2n+1$ is then
\begin{equation}
\frac{1}{\Box^{n}} \ \prd {\cal R}^{[n]}{}_{;\mu_1 \cdots
\mu_{2n+1}} = 0 \ , \label{geomodd}
\end{equation}
a natural generalization of the Maxwell equation. Notice that the
Maxwell and Einstein cases are the only ones when these geometric
equations are local. In all cases, however, the Fronsdal operators
provide local, albeit partly gauge fixed, forms for them. As
anticipated in the previous sections, these are the least singular
fully gauge invariant equations, while more singular forms can be
obtained combining eqs. (\ref{geomeven}) and (\ref{geomodd}) with
their traces, as in section 2.

\section{The issue of gauge fixing}

In the previous sections we have seen how the non-local geometric
equations for higher-spin bosons can be reduced to the usual local
forms using the traces $\Lambda'$ of the gauge parameters to reach
the ``Fronsdal gauges'' ${\cal H} = 0$. The resulting Fronsdal
equations, however, present a subtlety, since they involve {\it
unconstrained} fields $\phi$, whose double traces $\phi''$ do not
vanish identically. As we shall see, this subtlety requires a
slight revision of the standard gauge-fixing procedure and, as a
result, the field equations themselves set to zero the double
traces on shell, leaving again the right set of modes.

Whereas in the usual local formulation the de Donder gauge
condition
\begin{equation}
{\cal D} \ \equiv \ \prd \, \phi \ - \ \frac{1}{2} \ \pr \, \phi'
\ = \ 0
\end{equation}
is used to reduce the Fronsdal equation to the canonical form
\begin{equation}
\Box \, \phi \ = \ 0 \ ,
\end{equation}
the same procedure runs into a difficulty in this case. Indeed,
while the gauge variation of ${\cal D}$ under (\ref{gaugetransf})
is in general
\begin{equation}
\delta \, {\cal D} \ = \ \Box \, \Lambda \ - \ \pr^{\, 2} \
\Lambda' \ , \label{dnaive}
\end{equation}
once the trace of the gauge parameter has been used to set ${\cal
H}=0$, eq. (\ref{dnaive}) can only be consistent if the de Donder
gauge condition is modified in order that it be {\it identically
traceless}. While not automatic, since in our case $\phi''$ does
not vanish identically as in \cite{fronsdal}, this can be achieved
following once more an iterative procedure. Before dealing with
the general case, it is perhaps instructive to discuss explicitly
this issue for a spin-4 field, the first case where the problem
presents itself.

For a spin-4 field in $D$ dimensions, one can modify ${\cal D}$ by
the addition of a higher trace, to finally impose
\begin{equation}
{\cal D} \ + \ \Delta  \ \equiv \ \prd \, \phi \ - \ \frac{1}{2} \
\pr \, \phi' \ + \ \frac{1}{2(D+2)} \ \eta \ \pr \, \phi'' \ = \ 0
\ ,
\end{equation}
that is {\it identically traceless}. When used in the Fronsdal
equation, in momentum space this leads to
\begin{eqnarray}
p^{\; 2} \, \phi_{\mu\nu\rho\sigma} \ + \ \frac{1}{D+2} \, \left[
\, \eta_{\mu\nu} \, p_\rho \, p_\sigma \ + \ \cdots \, \right] \,
\phi'' \ = \ 0 \ , \label{dedonspinfour}
\end{eqnarray}
and this equation can only have non-trivial eigentensors if
$p^2=0$, since the second term does not contain any dimensionful
parameter. With a light-like momentum $p_+$, choosing for instance
$(\mu,\nu,\rho,\sigma)$ equal to $(+,-,+,+)$, it is then simple to
see that eq. (\ref{dedonspinfour})  sets $\phi''=0$. At this
point, the gauge condition reduces to the usual de Donder form,
with a field now constrained as in \cite{fronsdal}. The non-local
equation thus propagates correctly spin-4 polarizations, despite
the fact that a {\it traceless} gauge parameter, left over after
setting ${\cal H}=0$, is originally accompanied by an {\it
unconstrained} gauge field.

In general, for a spin-$s$ field $\phi$ one can define
\begin{equation}
\Delta \ = \ \sum_{k \geq 2} \ \left[ \, c_{k,1} \, \eta^{k-1} \,
\pr \, \phi^{[k]} \ + \   c_{k,2} \, \eta^{k} \, \prd \phi^{[k]}
\, \right] \ ,
\end{equation}
where the coefficients $c_{k,1}$, available for $k \leq \left[
\frac{s}{2} \right]$, and $c_{k,2}$, available $k \leq \left[
\frac{s-1}{2} \right]$, can be determined recursively requiring
that
\begin{eqnarray}
\Delta'  =  \sum_{k \geq 2} \ \biggl\{ \!\! & \eta^{k-1} & \!\!\!
\prd \phi^{[k]} \biggl[ \,  2\,  c_{k,1} \ + \ c_{k,2} \, k \,
\left( D + 2(s-k-2)\right) \, \biggr]
\nonumber \\
&+&\!\!\!\! c_{k,1} \, (k-1) \, \left( D + 2(s-k-1)\right)
\eta^{k-2} \, \pr \phi^{[k]}
\nonumber \\
&+& \!\!\! \! c_{k,1} \,\eta^{k-1} \, \pr \phi^{[k+1]} +
 c_{k,2} \, \eta^{k} \,
\prd \phi^{[k+1]}  \, \biggr\}
\end{eqnarray}
cancel the trace of the naive de Donder condition ${\cal D}$, {\it
i.e.} that
\begin{equation}
\Delta' \ - \ \frac{1}{2} \ \pr \; \phi'' \ = \ 0 \ ,
\end{equation}
so that
\begin{equation}
{\cal D} \ + \ \Delta \ = \ 0 \label{ddelta}
\end{equation}
is an allowed gauge condition. This requires that
\begin{equation}
c_{k+1,1} \ = \ - \ \frac{c_{k,1}}{k \, \left[ D+ 2(s-k-2)\right]}
\ ,
\end{equation}
and therefore
\begin{equation}
c_{k,1} \ = \ \frac{(-1)^k}{2\, (k-1)! \, \prod_{2\leq \ell \leq
k}\, \left[ D+ 2(s-\ell-1)\right]}   \qquad ( \, k \, \geq \, 2 \,
) \ . \label{ck1}
\end{equation}
The $c_{k,2}$ are then determined by the inhomogeneous difference
equation
\begin{equation}
c_{k+1,2} \ + \ \frac{c_{k,2}}{(k+1) \, \left[ D+ 2(s-k-3)\right]}
\ = \ g_k \ ,
\end{equation}
where
\begin{equation}
g_k \ = \ \frac{(-1)^k}{2\, (k+1)! \, \prod_{2\leq \ell \leq
(k+2)}\, \left[ D+ 2(s-\ell-1)\right]}  \ ,
\end{equation}
whose solution is
\begin{equation}
c_{k,2} \ = \ \frac{(-1)^{k-1}}{2\, (k-1)! \, \prod_{2\leq \ell
\leq (k+1)}\, \left[ D+ 2(s-\ell-1)\right]}   \qquad ( \, k \,
\geq \, 2 \, ) \ . \label{ck2}
\end{equation}

As a result, the field equations reduce in general to
\begin{equation}
\Box \, \phi + \pr \, \Delta \ = \ 0 \ , \label{boxdelta}
\end{equation}
while the existence of non-vanishing solutions requires that
$p^2=0$, the only covariant condition that can emerge from a
kinetic operator, invertible as a result of the gauge-fixing
procedure, that only involves dimensionless parameters. Eqs.
(\ref{ddelta}) then turn (\ref{boxdelta}) into
\begin{equation}
\pr \, {\cal D} \ = \ 0 \ ,
\end{equation}
and, with a light-like momentum $p_+$, it is easy to convince
oneself that these imply
\begin{equation}
{\cal D} \ = \ 0 \ ,
\end{equation}
and therefore, via their traces, $\phi''=0$.

When combined, as usual, with residual gauge transformations, the
de Donder conditions thus retrieved leave only the polarizations
corresponding to irreducible, traceless symmetric spin-$s$
tensors. Still, the need for these generalized $({\cal D} +
\Delta)$-gauges appears to reflect interesting properties of this
formulation at the quantum level, that deserve further
investigation.

\section{Higher-spin geometry and String Theory}

We now want to show that the higher-spin geometry discussed in the
previous sections plays a clear, albeit indirect, role in String
Theory. That this should be the case can be simply anticipated.
Following \cite{ko,witten}, the free equations of String Field
Theory can indeed be written in the form
\begin{equation}
{\cal Q} \ | \Phi \rangle \ = \ 0 \ , \label{eqstring}
\end{equation}
where ${\cal Q}$ denotes the BRST operator of the first-quantized
string, and have the chain of gauge invariances
\begin{equation}
\delta | \Phi^{(n)} \rangle \ = \ \ {\cal Q} \, | \Phi^{(n+1)}
\rangle \ , \label{gaugestring}
\end{equation}
with $\Phi^{(1)}$ the string gauge field and $\Phi^{(n)} \ ( n >
1)$ the corresponding chain of gauge (per gauge) parameters. It
should be appreciated that neither (\ref{eqstring}) nor
(\ref{gaugestring}) involve trace conditions. Whereas they
describe massive higher-spin modes, their limit as $\alpha' \to
\infty$ is thus expected to result in  fully gauge invariant free
massless higher-spin equations with no trace constraints like
those in \cite{fronsdal}. This is indeed the case, and for
symmetric tensors (say, in the ``leading Regge trajectory''
generated by the lowest string oscillators $\alpha_{-1}$ in the
bosonic string), the limit $\alpha' \to \infty$ yields the
equations
\begin{eqnarray}
&& \Box \, \phi \ = \ \partial \, C \ , \nonumber \\
&& \partial \, \cdot \, \phi \ - \ \partial \, D \ = \ C \ , \nonumber \\
&& \Box \, D \ = \ \partial \, \cdot \, C \ , \label{bosestring}
\end{eqnarray}
that link the {\it unconstrained} spin-$s$ tensors $\phi$,
spin-$(s-1)$ tensors $C$ and spin-$(s-2)$ tensors $D$, are
invariant under the {\it unconstrained} gauge transformations
\begin{eqnarray}
&& \delta \, \phi \ = \ \partial \, \Lambda \ , \nonumber \\
&& \delta \, C \ = \ \Box \, \Lambda \ , \nonumber \\
&& \delta \, D \ = \ \partial \cdot \Lambda \ ,
\label{bosestringauge}
\end{eqnarray}
and follow from the Lagrangians
\begin{eqnarray}
{\cal L} & = & - \ \frac{1}{2} \ \partial_\mu \, \phi \
\partial^\mu \, \phi
\ - \ \phi \ \partial \, C \ - \ s \ C \ \partial \, D \nonumber \\
&& - \ \frac{s}{2} \ C^2 \ - \ \left( {s \atop 2} \right) \
\partial_\mu \, D \
\partial^\mu \, D \ .
\end{eqnarray}
On can also eliminate the auxiliary field $C$, reducing eqs.
(\ref{bosestring}) to the pair of equations
\begin{eqnarray}
&&{\cal F} \ = \ \partial^2 \, \left( \varphi^{'} - 2D\right) \ , \nonumber \\
 && \Box \; D \ = \ \frac{1}{2} \, \partial \cdot \partial \cdot
 \varphi \ - \ \frac{1}{2} \, \partial \ \partial \cdot D \ ,
 \label{tripletnoB}
\end{eqnarray}

One can also introduce similar fermionic systems,
\begin{eqnarray}
&& \dsll \, \psi \ = \ \partial \, \chi \ , \nonumber \\
&& \partial \, \cdot \, \psi \ - \ \partial \, \lambda \ = \ \dsll
\, \chi \ ,
\nonumber \\
&& \dsll \lambda \ = \ \partial \, \cdot \, \chi \ ,
\label{fermistring}
\end{eqnarray}
that link the {\it unconstrained} spin-$(s+1/2)$ tensor-spinors
$\psi$, spin-$(s-1/2)$ tensor-spinors $\chi$ and spin-$(s-3/2)$
tensor-spinors $\lambda$, are invariant under the {\it
unconstrained} gauge transformations
\begin{eqnarray}
&& \delta \, \psi \ = \ \partial \, \epsilon \ , \nonumber \\
&& \delta \, \chi \ = \ \dsll \, \epsilon \ , \nonumber \\
&& \delta \, \lambda \ = \ \partial \cdot \epsilon \ ,
\end{eqnarray}
and follow from the rather simple Lagrangians
\begin{eqnarray}
{\cal L} & = & i \, \bar{\psi} \, \dsll \,\psi \ - \ i \,
\bar{\psi} \
\partial \chi \ + \ i \, \partial \, \bar{\chi} \ \psi  -
i \, s \, \bar{\chi} \, \dsll \, \chi \ \nonumber \\
&& + \ i \, s \, \bar{\chi} \,
\partial \, \lambda \ - \ i \, s \, \partial \bar{\lambda} \, \chi
\ - \ i\, s(s-1) \, \bar{\lambda} \, \dsll \, \lambda \ .
\end{eqnarray}

We recently came to know that actually the system
(\ref{bosestring}), in the slightly different form
(\ref{tripletnoB}, first emerged from a study of the lower
excitations of the bosonic string in the low-tension limit in the
work of A. Bengtsson \cite{abengt}, and played a key in the more
recent, interesting BRTS constructions of Pashnev, Tsulaia and
others \cite{pastsu}, that we shall comment further upon in the
next sections. On the other hand, the fermionic system
(\ref{fermistring}) does not play directly a role in the
superstring, where totally symmetric tensor-spinors are eliminated
by the GSO projection, but it does enter the spectrum of type-0
strings \cite{type0}. Generalizations of these equations to
tensors with mixed symmetry can also be obtained as interesting
applications of the BRST procedure, as in \cite{pastsu,sagtsu},
but for brevity here we shall confine our attention to the simpler
case of fully symmetric tensors.

It is important to notice that eqs. (\ref{bosestring}) and
(\ref{fermistring}), that we shall henceforth refer to as
``triplets'', propagate lower spins as well. This can be foreseen
rather simply in the light-cone description of the string states,
where, say, the physical polarizations for the system
(\ref{bosestring}) would correspond to states created by products
of transverse string oscillators of the type $\alpha_{-1}^{i_1}
\cdots \alpha_{-1}^{i_s}\, | 0 \rangle$, that indeed describe a
nested chain of modes of decreasing spins. The simplest
non-trivial case of this type, $\alpha_{-1}^{i_1}
\alpha_{-1}^{i_2}\, | 0 \rangle$, presents itself at the first
massive level of the open bosonic string. In the $\alpha' \to
\infty$ limit these states become massless, and while the
traceless part of this two-tensor describes massless spin-2 modes,
its trace describes an independent scalar mode. In general, one
thus expects that the local system (\ref{bosestring}) propagate
independent sets of irreducible modes of spins $s$, $s-2$, and so
on, down to zero or one according to whether $s$ is even or odd.

In order to analyze directly the physical polarizations described
by eqs. (\ref{bosestring}), it is convenient to choose the gauge
$C=0$, that can be reached solving a wave equation for the gauge
parameter $\Lambda$ and reduces the system to a pair of wave
equations for $\phi$ and $D$, together with the constraint
\begin{equation}
 \partial \, \cdot \, \phi \ - \ \partial \, D \ = \ 0 \ .
\label{bosefixstring}
\end{equation}
One can now take for $\phi$ and $D$ two plane waves characterized
by the same light-like momentum $p_+$, and the result follows if
the constraints of eq. (\ref{bosefixstring}) are combined with
residual gauge transformations also described by plane waves of
momentum $p_+$, that as such do not affect the original gauge
choice $C=0$, obtaining
\begin{eqnarray}
&& \delta \phi_{(+)^k \, (a)^{s-k}} \ = \ k \ p_+ \,
\Lambda_{(+)^{k-1} \, (a)^{s-k}} \ , \label{gaugevstring} \\
&&  p_+ \, \phi_{- \, (+)^k \, (a)^{s-1-k}} \ = \ k \ p_+  \,
D_{(+)^{k-1} \, (a)^{s-1-k}}  \ , \label{dedondstring}
\end{eqnarray}
where $a$ stands collectively for any of the other directions,
``-'' or ``i''. Using eq. (\ref{gaugevstring}), one can eliminate
all components of the gauge field $\phi$ with at least a lower
``+'' index, and then eq. (\ref{dedondstring}) kills all other
components with no lower ``+'' indices and at least a lower ``-''
index, while the other choices of $k$ in eq. (\ref{dedondstring})
dispose of the whole of $D$. In conclusion, one is left with the
transverse components $\phi_{i_1 \cdots i_s}$, that indeed
describe the (traceless) polarizations of a spin $s$ field,
together will all their traces, that indeed correspond to the
chain of spins $s$, $s-2$, and so on, down to zero or one
according to whether $s$ is even or odd. A similar, if more
involved, argument, applies to the fermionic system, and shows
that, with a spinor-tensor $\psi_{\mu_1 \cdots \mu_n}$ and its
$\chi$ and $\lambda$ companions, one is actually propagating
\emph{all} half-integer lower spins, together with the maximum
one.

Confining for simplicity our attention to the bosonic system
(\ref{bosestring}), we can now show how, after a suitable
elimination of the $C$ and $D$ fields, $\phi$ satisfies the
non-local equations of \cite{fs}. This can be done in two
different ways. The first consists in the direct elimination of C
from the divergence of the second of eqs. (\ref{bosestring}), that
when combined with the third gives
\begin{equation}
\Box \, D \ + \ \frac{1}{2} \, \pr \, \prd D \ = \ \frac{1}{2} \,
\prd \prd \phi \ . \label{prdprd}
\end{equation}
Inverting the differential operator in (\ref{prdprd}) defines a
solution for $D$ that, by construction, eliminates all lower-spin
modes, and substituting it in the second of (\ref{bosestring})
expresses $C$ in terms of $\phi$. Finally, substituting $C$ in the
first of eqs. (\ref{bosestring}) yields a non-local fully gauge
invariant equation equivalent to (\ref{fd3h}). This can be seen
explicitly for the first few cases, but in general one can find a
simpler argument noticing that $\phi$ and $D$ define the chain of
lower-spin fields $(n \geq 1)$
\begin{equation}
{\cal W}_{n} \ = \ \phi^{\, [n]}  \ - \ 2 \, n \,  D^{\, [n-1]} \
 \end{equation} Eqs. (\ref{bosestring}) and
(\ref{bosestringauge}) imply that these fields satisfy the chain
of inhomogeneous Fronsdal equations
\begin{equation}
{\cal F} \left( {\cal W}_n \right) \ = \ \pr^{\, 2} \  {\cal
W}_{n+1} \ ,
\end{equation}
and transform under gauge transformations as ordinary fields of
spin $s - 2n$, {\it i.e.}
\begin{equation}
\delta \ {\cal W}_n \ = \ \pr \, \Lambda^{[n]} \ .
\end{equation}

One can now look for solutions of eqs. (\ref{bosestring}) that are
free of lower-spin excitations. Naively, this would correspond to
eliminating ${\cal W}_1$, but more naturally one can demand that
${\cal W}_1$ be pure gauge, letting
\begin{equation}
{\cal W}_1 \ = \ \phi' \ - \ 2\,  D \ = \ \pr \, \alpha \ ,
\label{trunc}
\end{equation}
where eq. (\ref{bosestring}) implies that a gauge transformation
affects $\alpha$ according to
\begin{equation}
\delta \, \alpha  \ = \ \Lambda' \ .
\end{equation}
Substituting in eqs. (\ref{bosestring}) then gives the
inhomogeneous Fronsdal equation
\begin{equation}
{\cal F} \left( \phi \right) \ = \ 3 \, \pr^3 \ \alpha \ ,
\label{froncomp}
\end{equation}
a local form in terms of the source $\alpha$ of the non-local
geometric equations of \cite{fs}, as advertised. These can be
recovered deriving from (\ref{froncomp}) corresponding equations
for the ${\cal F}^{(n)}$,
\begin{equation}
{\cal F}^{(n)} \ = \ (2 \, n \ + \ 1) \ \frac{\pr^{\, 2
n+1}}{\Box^{n}} \, \alpha^{[n]} \ ,
\end{equation}
where by construction $\alpha$  is replaced by its multiple
traces, and thus disappears altogether for $n$ large enough, so
that eventually one retrieves the non-local homogeneous equations
of section 2. On the other hand, the direct elimination of ${\cal
W}_1$ leads to the usual homogeneous Fronsdal equation with a
traceless gauge parameter, but, as is section 5, with an
unconstrained gauge potential.

\section{Local compensator forms of the field equations in flat space}

In the previous section we have seen how a single spin-$s$ field
can be described by a ``triplet'' whose lower-spin excitations are
restricted to be pure gauge by eq. (\ref{trunc}), a choice
motivated by the fact that $\varphi^{'} - 2 D$ transforms as a
canonical spin-$(s-2)$ field. This indeed turns the first of eqs.
(\ref{tripletnoB}) into (\ref{froncomp}), where $\alpha$ is a
spin-$(s-3)$ field that plays the role of the single compensator
needed in this formulation, but the second eq. (\ref{tripletnoB})
takes an apparently more complicated form, and becomes
\begin{equation}
\Box \, \varphi^{'} \ + \ \frac{1}{2}\, \partial \ \partial \cdot
\varphi^{'} \ - \ \partial \cdot \partial \cdot \varphi \ = \
\frac{3}{2} \, \Box \, \partial \alpha \ + \ \partial^2 \
\partial \cdot \alpha \ . \label{deq}
\end{equation}

In terms of the Fronsdal operator, however, this simplifies
considerably, since eq. (\ref{fronsdeq}) indeed implies that
\begin{equation}
{\cal F}^{'} \ = \ 2 \, \Box \, \varphi^{'} \ - \ 2\, \partial
\cdot
\partial \cdot \varphi \ + \ \partial \ \partial \cdot \varphi^{'} \ + \ \partial^2 \
\varphi^{''}\ ,
\end{equation}
so that (\ref{deq}) is equivalent to
\begin{equation}
{\cal F}^{'} \ - \ \partial^2 \, \varphi^{''} \ = \ 3\, \Box
\partial \alpha \ + \ 2 \, \partial^2 \ \partial \cdot \alpha \ .
\end{equation}
On the other hand, the trace of eq. (\ref{froncomp}) is
\begin{equation}
{\cal F}^{'} \ = \ 3\, \Box \partial \alpha \ + \ 6 \,
\partial^2 \ \partial \cdot \alpha \ + \ 3\,
\partial^3 \alpha^{'} \ ,
\end{equation}
and thus, by comparison, one obtains
\begin{equation}
\partial^2 \, \varphi^{''} \ = \ 4 \, \partial^2 \ \partial \cdot \alpha \
+ \ \partial^3 \ \alpha^{'} \ = \ \partial^2\ \left( 4
\partial \cdot \alpha \ + \ \partial \alpha^{'} \right) \ .
\end{equation}

The conclusion is that the ``triplet equations''
(\ref{bosestring}) imply a pair of \emph{local} equations for a
single massless spin-$s$ gauge field $\varphi$ and a single
spin-$(s-3)$ compensator $\alpha$,
\begin{eqnarray}
&& {\cal F} \ = \ 3\, \partial^{\; 3} \, \alpha \ , \nonumber \\
&& \varphi^{''} \ = \ 4\,  \partial \cdot \alpha \ + \ \partial \,
\alpha^{'} \ , \label{flatbosecompens}
\end{eqnarray}
that are invariant under the \emph{unconstrained} gauge
transformations
\begin{eqnarray}
&& \delta \; \varphi \ = \ \partial \; \Lambda \ , \\
&& \delta \; \alpha \ = \ \Lambda^{'} \ ,
\end{eqnarray}
and clearly reduce to the standard Fronsdal form after a partial
gauge fixing using the trace $\Lambda^{'}$ of the gauge parameter.

These local equations are nicely consistent, since the second is
implied by the first, as can be seen using the Bianchi identity of
eq. (\ref{bianchifron}). However, it is important to stress that
\emph{these are not Lagrangian equations}, somewhat in the spirit
of the Vasiliev form of higher-spin dynamics \cite{fvas,vaseq}. A
Lagrangian form for the compensator equations can be obtained, as
in \cite{pastsu,sagtsu}, at
the expense of introducing a number of additional fields, although
the resulting procedure is nicely guided by the BRST approach.

One can also repeat the construction of local compensator
equations for Fermi fields. Here the story parallels the
discussion in subsection 3, since the fermionic Fang-Fronsdal operator
(\ref{fangfroneq}) varies into a term proportional to the
gamma-trace of the gauge parameter,
\begin{equation}
\delta {\cal S} \ = \ - 2 \, i \ \partial^2 \esl \ ,
\end{equation}
under the gauge transformation
\begin{equation}
\delta \psi \ = \ \partial \; \epsilon \ .
\end{equation}
In addition, ${\cal S}$ satisfies the ``anomalous'' Bianchi
identity (\ref{fangfronbianchi}), and as a result the gauge
parameter and the gauge field were constrained in \cite{fronsdal}
to satisfy the conditions
\begin{equation}
\esl \ = \ 0 \ , \qquad \ \psisl\;' \ = \ 0 \ .
\end{equation}

As for integer-spin fields, one can also eliminate these
constraints introducing a single compensator field $\xi$. The
resulting equations,
\begin{eqnarray}
&& {\cal S} \ = \ - \ 2 \, i \, \partial^2 \, \xi \ , \nonumber \\
&& \psisl^{\ '} \ = \ 2 \, \partial \cdot \xi \ + \ \partial \,
\xi^{\ '} \ + \ \dsll \xisl \ , \label{compfermiflat}
\end{eqnarray}
are then invariant under the gauge transformations
\begin{eqnarray}
&& \delta \psi \ = \ \partial \, \epsilon \ , \nonumber \\
&& \delta \xi \ = \ \esl \ ,
\end{eqnarray}
involving an unconstrained gauge parameter, and are nicely
consistent, since the first implies the second via the Bianchi
identity (\ref{fangfronbianchi}). As in the bosonic case, however,
these are not Lagrangian equations. Corresponding Lagrangian
equations have not appeared yet, but they are expected to follow from
a direct, if tedious, procedure, along the lines of
\cite{pastsu}.

\section{Local compensator forms of the field equations in (A)dS}

One can also extend the spin-$s$ compensator
equations of the previous section to (anti)de Sitter
backgrounds, henceforth denoted by (A)dS for brevity. This was done
in \cite{sagtsu}, and it is worthwhile to review briefly these
results, given also the central role of (A)dS backgrounds in the Vasiliev
formulation \cite{vaseq}. While \cite{pastsu} contains also
derivations of Lagrangian equations based on the BRST formalism, here we
shall simply content ourselves with an account of the direct
constructions of \cite{sagtsu}.

To this end, one needs only two additional inputs,
the commutator of two covariant derivatives on a vector
\begin{equation}
[ \nabla_\mu ,
\nabla_\nu ] \, V_\rho \ = \ \frac{1}{L^2} \left( g_{\nu\rho} \,
V_\mu \ - \
 g_{\mu\rho} \, V_\nu \right) \ , \label{noncommvect}
\end{equation}
where $L$ determines the AdS curvature and the dS case is recovered formally continuing $L$ to imaginary values,
and the corresponding commutator on a spinor,
\begin{equation}
[ \nabla_\mu , \nabla_\nu ] \, \eta \ = \ - \ \frac{1}{2 L^2} \
\gamma_{\mu\nu} \ \eta \ , \label{noncommspin}
\end{equation}
where $\gamma_{\mu\nu}$ is antisymmetric in $\mu$ and $\nu$ and
equals the product $\gamma_\mu \gamma_\nu$ when $\mu$ and $\nu$
are different, since these determine all other cases.

Let us therefore begin by considering the bosonic case, noticing that
the gauge transformations for the fields $\varphi$ and $\alpha$ in
such a curved background take naturally the form
\begin{eqnarray}
&& \delta \; \varphi \ = \ \nabla \Lambda \ , \nonumber \\
&& \delta \; \alpha \ = \ \Lambda^{'} \ ,
\end{eqnarray}
where $\nabla$ denotes and (A)dS covariant derivative.

One can then proceed in various ways, for instance starting from
the gauge variation of the (A)dS form of the Fronsdal operator
\begin{eqnarray}
\delta {\cal F}_L &\equiv& \delta \; \left\{ {\cal F} -
\frac{1}{L^2}\,
[(3-{\cal D}-s)(2-s) -s] \varphi - 2 g \varphi^{'} \right\} \nonumber \\
&=& 3 \; ( \nabla^3 \Lambda^{'}) \, - \, \frac{4}{L^2}\ g \nabla
\Lambda^{'} \ ,
\end{eqnarray}
so that
\begin{equation}
{\cal F}_L \ = \ {\cal F} \ - \ \frac{1}{L^2} \, \left\{ \left[
(3-{\cal D}-s)(2-s) - s \right]\, \varphi \ + \ 2 \, g \;
\varphi^{'} \right\} \ ,
\end{equation}
and it is then simple to conclude that the compensator form of the
higher-spin equations in (A)dS is
\begin{eqnarray}
&& {\cal F} \ = \ 3 \nabla^3 \alpha + \frac{1}{L^2}\ \{ [(3-{\cal
D}-s)(2-s) -s] \varphi + 2 g \varphi^{'} \} - \frac{4}{L^2}\ g
\nabla
\alpha \ , \nonumber \\
&& \varphi^{''} \ = \ 4 \nabla \cdot \alpha \ + \ \nabla
\alpha^{'} \ . \label{AdScompens}
\end{eqnarray}

These are again nicely consistent, as can be seen making use of the Bianchi
identity, that now becomes
\begin{equation}
\nabla \cdot {\cal F}_L \ - \ \frac{1}{2} \, \nabla \, {\cal
F}_L^{'} \ = \ - \ \frac{3}{2} \ \nabla^3 \ \varphi^{''} \ + \
\frac{2}{L^2} \ g \ \nabla\, \varphi^{''} \ , \label{bianchiads}
\end{equation}
and one can in fact verify that the
first of (\ref{AdScompens}) implies the second.

The fermionic compensator equations (\ref{compfermiflat})
can be also generalized to (A)dS backgrounds. The gauge transformation for a spin-$s$ fermion
becomes in this case
\begin{equation}
\delta \psi \ = \ \nabla \, \epsilon \ + \ \frac{1}{2 L} \, \gamma
\; \epsilon \ ,
\end{equation}
where $\nabla$ denotes again an (A)dS
covariant derivative, and in
order to proceed, one needs only the commutator of two covariant
derivatives on a spin-1/2 field of eq. (\ref{noncommspin}).

For a spin-$s$ fermion ($s=n+\frac{1}{2}$), where $n$ is the
number of vector indices carried by the field $\psi$, the
compensator equations in an (A)dS background are then
\begin{eqnarray}
&& \left( \nablasl \, \psi \ - \ \nabla \psisl \right) \ + \
\frac{1}{2 L} \, \left[ {\cal D}\ + \ 2( n \, - \, 2) \right] \psi
\ + \ \frac{1}{2L} \, \gamma \, \psisl \nonumber \\
&& \qquad\qquad\qquad\quad = \ -\,  \{ \nabla, \nabla \} \xi \ + \
\frac{1}{L} \, \gamma \, \nabla \, \xi \ + \
\frac{3}{2 L^2} \, g \, \xi \ , \nonumber \\
&& \psisl^{\ '} \ = \ 2 \, \nabla \cdot \xi \ + \ \nablasl \xisl \
+ \ \nabla \xi^{\ '} \ + \ \frac{1}{2 L} \, \left[ {\cal D}\ + \
2( n \, - \, 2) \right] \, \xisl  \ - \ \frac{1}{2L} \, \gamma \,
\xi^{\ '} \ , \label{fermiadsnl}
\end{eqnarray}
and are invariant under
\begin{eqnarray}
&& \delta \psi \ = \ \nabla \, \epsilon \ , \nonumber \\
&& \delta \xi \ = \ \esl \ ,
\end{eqnarray}
with an unconstrained parameter $\epsilon$.

Eqs. (\ref{fermiadsnl}) are again non-Lagrangian,
like their flat-space counterparts (\ref{compfermiflat}), but are
again nicely consistent, on account of the (A)dS deformation of
the Bianchi identity (\ref{bianchifermi}),
\begin{eqnarray}
\nabla \cdot {\cal S} \ - \ \frac{1}{2} \, \nabla \, {\cal S}{\;
'} \ - \ \frac{1}{2} \, \nablasl  \ssl &=& \frac{i}{4L } \, \gamma
\, S^{\, \prime} \ + \ \frac{i}{4L} \left[ ({\cal D} \, - \, 2)\,
+\, 2\; (n\, - \, 1) \right] \, \ssl \nonumber \\ &+& \
 \frac{i}{2} \ \left[  \{ \nabla , \nabla \} \ - \ \frac{1}{L}\ \gamma \, \nabla
\ - \ \frac{3}{2\;L^2}\right] \, \psisl\;'
 \ ,
\end{eqnarray}
where now the Fang-Fronsdal operator ${\cal S}$ is also deformed
and becomes
\begin{equation}
{\cal S} \ = \ i\, \left( \nablasl \, \psi \ - \ \nabla \psisl
\right) \ + \ \frac{i}{2 L} \, \left[ {\cal D}\ + \ 2( n \, - \,
2) \right] \psi \ + \ \frac{i}{2L} \, \gamma \, \psisl \ .
\end{equation}

\section{Conclusions}

It is becoming increasingly clear that a deeper understanding of
String Theory cannot forego a closer look at its massive
excitations. As a result, Higher Spin Gauge Theory is quickly
coming to the forefront of current research, since it appears to
provide key examples of dynamical systems whose structural
complexity is somehow intermediate between those of String Theory
and of ordinary low-spin theories. With this is mind, in \cite{fs}
we reconsidered the standard free higher-spin equations,
exhibiting a geometric structure that makes them more akin to
their low-spin counterparts. This rests on the replacement, for
spin greater than two, of the ordinary local kinetic operators of
\cite{fronsdal} with non-local ones that guarantee the invariance
of the field equations under gauge transformations with
unconstrained parameters and the absence in the Bianchi identities
of ``anomalous'' terms depending on the double traces (or triple
$\gamma$-traces) of the gauge fields. The end result can be linked
to the generalized curvatures introduced by de Wit and Freedman
\cite{dewf}, as the local Fronsdal equations were linked in
\cite{dewf} to the traces of the second-order connections.
Suitable compensators can dispose of the non-local terms, while
preserving the unconstrained gauge invariance, but only at the
price of making the geometry that underlies the equations less
transparent, and therefore they were only briefly discussed in
\cite{fs}.

In \cite{fs} we had also noticed that the geometric equations can
always be turned to the local, Fronsdal form after partial gauge
fixings involving the traces of the gauge parameters. Here we have
supplemented this result, showing that if the usual de Donder
gauge conditions are modified in order that they be {\it
identically traceless}, as demanded by the presence of {\it
unconstrained fields}, the resulting field equations set to zero
their double traces on shell. Once these are removed, the study of
on-shell polarizations proceeds as in the standard case, despite
the apparent initial presence of more field components together
with the same number of traceless gauge parameters.

Finally, we have shown how String Theory makes a full, if
indirect, use, of the curvatures of \cite{dewf}. In the bosonic
case, these are embodied in a set of local gauge invariant
equations propagating modes of spin $s$, $s-2$, and so on, down to
zero or one according to whether $s$ is even or odd, that follow
from the Stueckelberg-like equations for massive string modes in
the limit $\alpha' \to \infty$. As we have seen, these reduce to
the non-local equations of \cite{fs} once one restricts the
attention to fields that are free of lower-spin excitations.

The string-inspired systems of ``triplets'', however, are of
interest in their own right, can be simply extended \cite{sagtsu} to
the mixed-symmetry tensors discussed in \cite{bekaert}, and their
ability to describe nested chains of lower spins may well prove
valuable in establishing more direct links between String Theory
and Higher-Spin Gauge Theory, or in providing level-like
truncations for the latter. For the time being, following
\cite{sagtsu}, we have seen how the ``triplet'' equations of
section 6 provide an interesting route toward simple
non-Lagrangian local compensator forms of the field equations for
higher-spin bosons and fermions, that we have displayed both for a
flat and an (A)dS background, that plays a key role in the
Vasiliev formulation \cite{vaseq}. These display the same
unconstrained gauge symmetry as the non-local equations of the
previous sections, and are also nicely determined by the BRST
approach of \cite{pastsu,bpsz}. More details, together with
local Lagrangian forms for the bosonic equations, are discussed at length in \cite{sagtsu}.

It will be very interesting to elucidate the possible role of the
higher-spin geometry reviewed here in the Vasiliev formulation \cite{vaseq}, or
in the conformal higher-spin theory proposed by Segal in
\cite{segal}, also in view of its relation to String Theory and to the BRST-like systems studied
in \cite{buch,pastsu}. We hope to return to some of these issues in
the near future.

\acknowledgments

It is a pleasure to thank M. Bianchi, G. Pradisi and especially A.
Segal for stimulating conversations. This work was supported in
part by I.N.F.N., by the EC contract HPRN-CT-2000-00122, by the EC
contract HPRN-CT-2000-00148, by the INTAS contracts 99-1-590 and
03-51-6346, by the MURST-COFIN contracts 2001-025492 and
2003-023852, and by the NATO Contract PST.CLG.978785. Finally,
A.S. would like to thank M. Tsulaia for a stimulating
collaboration on the topics reviewed in the last sections.

\end{document}